\documentclass[aps,prl,twocolumn,amsmath,amssymb,nofootinbib,showpacs,superscriptaddress]{revtex4-2}

\usepackage{times}

\usepackage{color} 
\usepackage{graphics} 

\usepackage{graphicx}
\usepackage{dcolumn}
\usepackage{bm}

\newcommand{\rednew}[1]{#1}

\begin{document}


\title{Super-resolution linear optical imaging in the far field}

\author{A.A. Pushkina}
\thanks{These authors contributed equally to this work.}
\affiliation{Department of Physics, University of Oxford, Oxford, OX1 3PU, UK}
\author{G. Maltese}
\thanks{These authors contributed equally to this work.}
\affiliation{Department of Physics, University of Oxford, Oxford, OX1 3PU, UK}
\author{J.I. Costa-Filho}
\thanks{These authors contributed equally to this work.}
\affiliation{Department of Physics, University of Oxford, Oxford, OX1 3PU, UK}
\author{P. Patel}
\affiliation{Department of Physics, University of Oxford, Oxford, OX1 3PU, UK}
\author{A. I. Lvovsky}
\email{Alex.Lvovsky@physics.ox.ac.uk}
\affiliation{Department of Physics, University of Oxford, Oxford, OX1 3PU, UK}
\affiliation{Russian Quantum Center, 100 Novaya St., Skolkovo, Moscow, 143025, Russia}
\affiliation{P. N. Lebedev Physics Institute, Leninskiy prospect 53, Moscow, 119991, Russia}


\begin{abstract}
The resolution of optical imaging devices is ultimately limited by the diffraction of light. To circumvent this limit, modern super-resolution microscopy techniques employ active interaction with the object by exploiting its optical nonlinearities, nonclassical properties of the illumination beam, or near-field probing.  Thus, they are not applicable whenever such interaction is not possible, for example, in astronomy or non-invasive biological imaging. Far-field, linear-optical super-resolution techniques based on passive analysis of light coming from the object
would cover these gaps. In this paper, we present the first proof-of-principle demonstration of such a technique.  It works by accessing information about spatial correlations of the image optical field and, hence, about the object itself via measuring projections onto Hermite-Gaussian transverse spatial modes. With a basis of 21 spatial modes in both transverse dimensions, we perform two-dimensional imaging  with twofold resolution enhancement beyond the diffraction limit.
\end{abstract}

\maketitle

The quest for improving resolution in optical imaging has always stumbled upon a seemingly unbreakable wall: the diffraction limit. The light field from the object, as it propagates through the imaging system, experiences diffraction, which gives rise to the smearing of the image. The diffraction limit is usually defined in terms of the heuristic Rayleigh's criterion $\theta = 1.22\lambda / D$, where $\theta$ is the resolvable angular separation, $\lambda$ the wavelength of light and $D$ the diameter of the objective lens' aperture \cite{rayleigh1879optics,pike2016resolution}. It bounds the resolution of optical microscopes to around 200 nm.

The diffraction limit is valid when objects are illuminated by classical light, the image is acquired in the far field, and the involved imaging processes are linear \cite{mandel1995optics,goodman2005fourier}. In the last decades we have witnessed an explosion of the so called super-resolution techniques, which could surpass the diffraction limit by breaking at least one of the aforementioned assumptions. By using e.g.~nonlinear excitation of fluorophores \cite{hell1994sted,hell2007nano}, utilizing their distinguishability in  time  \cite{rust2006storm,dickson1997switch}, or near-field probing of evanescent waves \cite{durig1986nearfield}, they were able to get around the diffraction barrier and bring optical microscopy to the nanoscale. 

Each of these methods however requires direct interaction with the sample and/or certain nonlinear properties thereof, and hence comes  with a host of limitations, in addition to significant cost and complexity. In certain cases, such as astronomical imaging or microscopy of certain sensitive samples \cite{tsang2019starlight}, the active nature of the interaction with the object precludes the application of existing superresolution techniques altogether.

A recent breakthrough \cite{tsang2016superresolution}, however, has shown that super-resolution can be achieved in the far field, with linear optics, and for standard illumination. It hinges on the discovery that the optical field's spatial correlations, which are ignored in conventional direct intensity measurements, contain additional information about the object. That information can be accessed by coherently processing the image field just before its detection, and therefore does not require any active manipulation of the sample.

One way to carry out this coherent processing is spatial mode sorting or demultiplexing \cite{tsang2016superresolution} of the image field, i.e., decomposing it into a basis of spatial modes, e.g., the Hermite-Gaussian (HG) basis \cite{tsang2016superresolution,rehacek2017beyond}, and measuring the magnitude of each component. The shape of the object is then reconstructed from these measurements. Our approach is therefore reminiscent of the so-called single-pixel imaging  \cite{duarte2008singlepixel, edgar2019singlepixel}, which utilizes spatial-mode decomposition where array detectors are unavailable because of wavelength restrictions or rapid acquisition requirements.

Intuitively, our approach helps achieving superresolution by leveraging the fine spatial structure of these modes: since the size of their features scales with the inverse square root of the mode order, measuring the image field's projections into higher-order modes accesses increasingly finer details of the spatial distribution of the field correlations and, therefore, the sub-wavelength information they carry.


The original theoretical idea \cite{tsang2016superresolution}, as well as all existing experimental work \cite{yang2016far,tsang2016fault,paur2016achieving,tham2017phase, parniak2018interference} focus on estimating just one or several parameters of the object, such as the separation  between two point sources. Here we report the first experimental demonstration of  this approach in application to full two-dimensional imaging. Our work is based on the method proposed theoretically by Yang {\it et al.} \cite{yang2016far} and dubbed Hermite-Gaussian microscopy (HGM). Tsang has shown the advantage of this approach in comparison with direct imaging in terms of quantum Fisher-information formalism \cite{tsang2017subdiffraction,tsang2019subdiffraction_quantum}. In practice, we are able to resolve the objects' details at one half of the diffraction limit imposed by the optical system.

Implementing HGM experimentally faces a number of challenges. First, the image reconstruction method of Ref.~\cite{yang2016far} assumes an ideal optical system with no aberrations and a perfectly Gaussian point spread function. Second, it requires precise measurement of the spatial mode in the HG basis which must be perfectly matched to the image field. This requirement is further complicated by the signal magnitudes associated with higher- and lower-order modes differing by several orders of magnitude. \rednew{Third, the method of Ref.~\cite{yang2016far} breaks down in the presence of noise; particularly it is  vulnerable to the shot noise inherent to the quantum nature of light (see Supplementary Materials).}

We overcome these challenges as follows. Instead of implementing direct  HG mode demultiplexing of the image field, 
we perform sequential heterodyne detection of that field with different HG beams as local oscillators (LO). The heterodyne detector is sensitive only to the component of the field that matches the LO mode, effectively selecting the necessary field projections. Since it is technically easier to prepare HG modes \cite{pushkina2020slm,forbes2016creation} than sort them \cite{zhou2018hermite,hiekkamaki2019perfect,fontaine2018sorter}, this approach significantly simplifies the experiment. We acquire heterodyne photocurrents for $HG_{m,n}$ with $m$ and $n$ ranging from 0 to 20. 

We overcome the imperfections of the imaging system and the systematic errors by using machine learning, rather than a rigid theoretical model, to calibrate our measurements. We train a neural network (NN) to reconstruct a superresolved image from the measurement data by presenting it with these data for a variety of known objects.
This training enables the NN to reconstruct an unknown object from a set of measurements acquired with the same setup. 

\begin{figure}[h!]
	\includegraphics[width=\columnwidth]{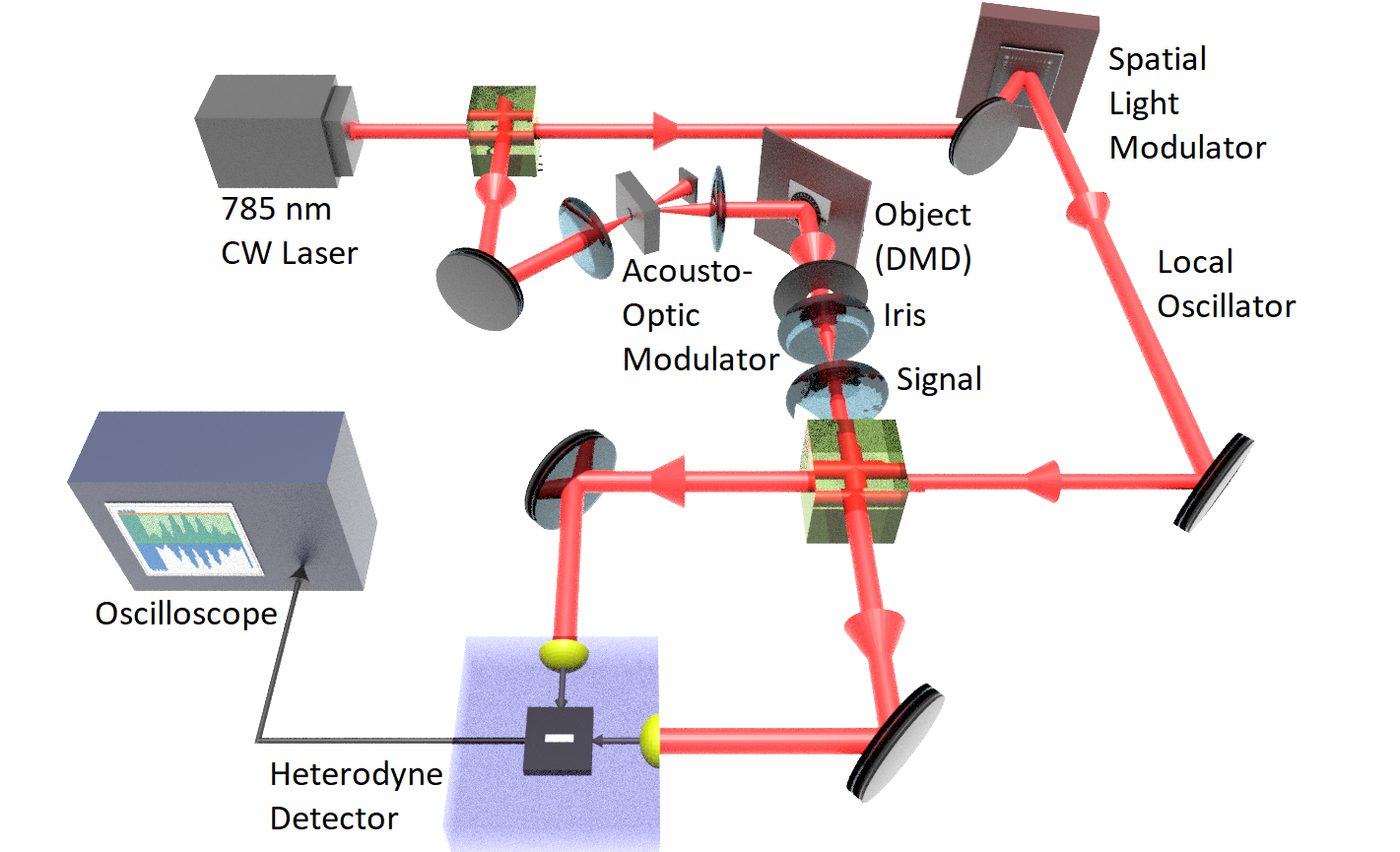}
	\caption{Schematic of the experiment. The object is a 210$\times$210 pixel bitmap displayed on the DMD; the spatial modes of the local oscillator are prepared using a liquid-crystal SLM.}
	\label{fig:setup}
\end{figure}

\begin{figure*}[t]
	\includegraphics[width=1.6\columnwidth]{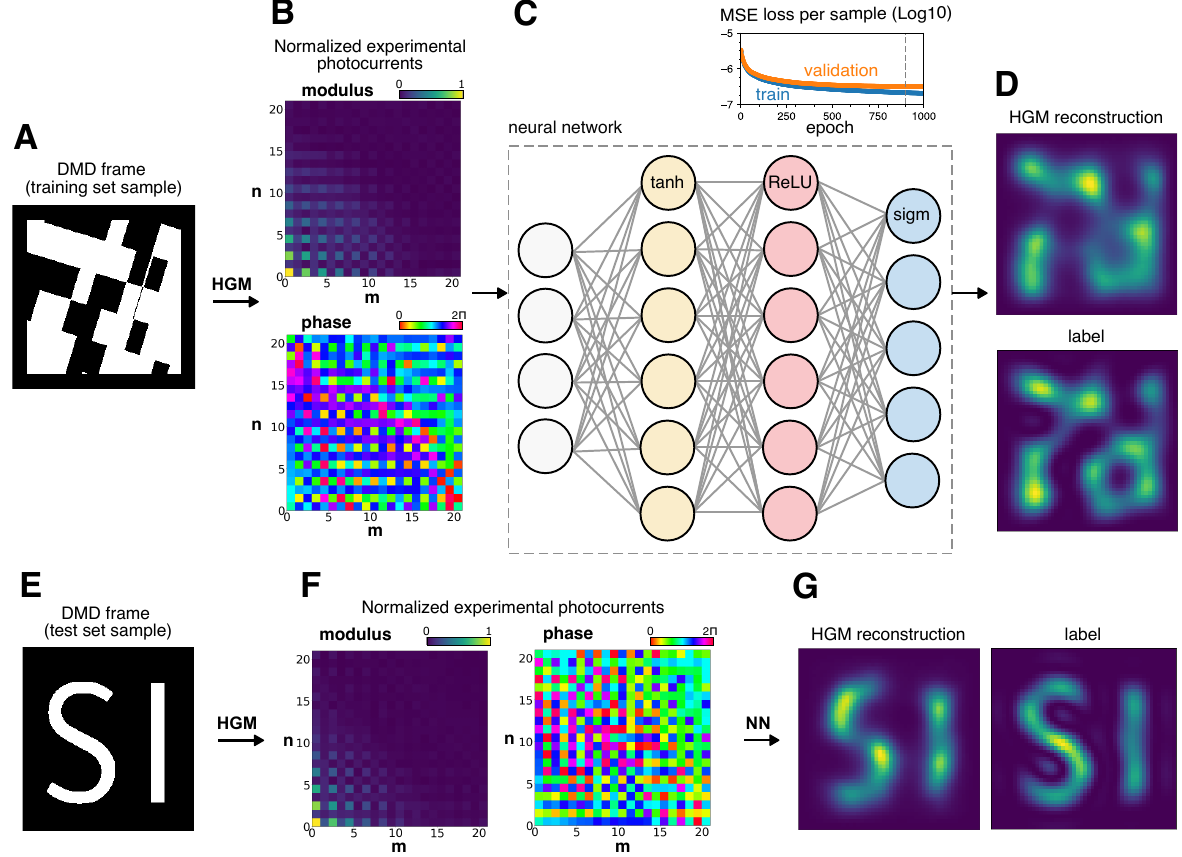}
	\caption{Sketch of the data acquisition and processing pipeline. We image via HGM a sample (A) and acquire its complex photocurrents (B), which are then fed to the NN (C) (the inset shows the learning curve, i.e. the training and cross-validation loss versus the training epoch). The NN is trained to predict the  idealized HGM reconstruction of the sample image (D). (E-G) show the reconstruction of a sample from the test set.
	}
	\label{fig:NN}
\end{figure*}

This proof-of-principle experiment is implemented in a simplified setting with coherently illuminated samples and an imaging system with a low numerical aperture. A schematic of the setup is presented in Fig.~\ref{fig:setup}. A continuous laser beam at 785 nm is initially split into the signal and LO paths. At the signal path, the beam is frequency-shifted by 92.05 MHz via an acousto-optic modulator (AOM) before illuminating a binary amplitude mask (the object to image) displayed via a digital micromirror device (DMD). The objects are generated inside a 210$\times$210 pixel area, with a pixel pitch of 7.56 $\mu \textup{m}$. To impose the diffraction limit, the light reflected from the DMD is imaged by an objective lens placed at a distance of 245.5 cm and with a numerical aperture (NA) reduced to $7.1 \times 10^{-4}$ by an iris of 3.5 mm diameter placed in front of it. The corresponding (theoretical) coherent light Rayleigh limit $1.64 \lambda/ 2 \textup{NA}$ is now 906 $\mu \textup{m}$ (120 DMD pixels) and, for comparison, the classical incoherent light limit $1.22 \lambda/ 2 \textup{NA}$ is 674 $\mu \textup{m}$ (89 DMD pixels). In the LO path, the laser beam is shaped into a $HG_{m,n}$ mode by a liquid-crystal spatial light modulator (SLM). Using the scheme of Ref.~\cite{bolduc2013exact}, which allows independent phase and amplitude modulation of the beam, plus a previously developed procedure \cite{pushkina2020slm} to compensate for the SLM's imperfections, we are able to generate high-quality HG modes up to the $20$-th order in both directions. Finally, the signal and LO paths are recombined for heterodyne detection. The produced photocurrents are demodulated; their phases and amplitudes are recorded (see Supplementary Materials).

\rednew{The signal-to-noise ratio of the acquired signal was limited by technical noise and amounted, dependent on the object, to 25--35 dB for lowest-order modes and 0--10 dB for highest-order modes. This ratio was typically poorer for odd modes because most objects had a dominant symmetric component, which gave rise to higher signal in even modes [Fig.~2(b)].}

\begin{figure}[b]
	\includegraphics[width=1\columnwidth]{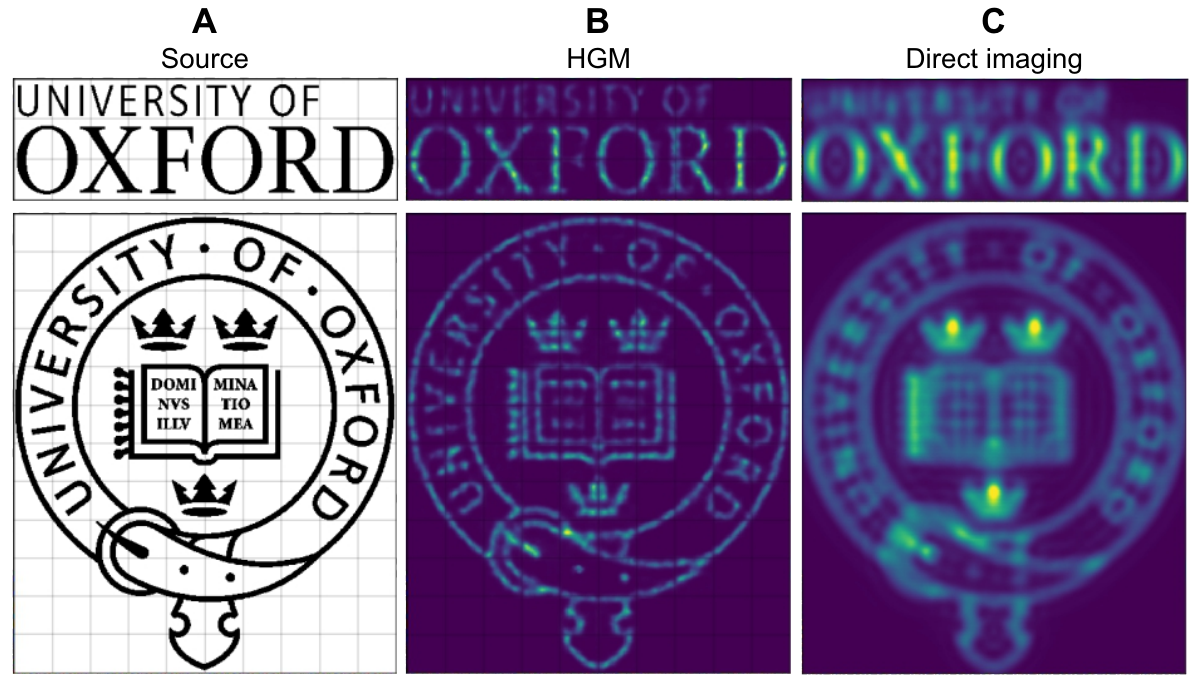}
	\caption{Original (A), HGM-reconstructed (B) and camera (C) images of the  Oxford logo and coat-of-arms test sets. The grid shows the division of the images into square sub-images that are fed to HGM sequentially.}
	\label{fig:reconstructions}
\end{figure}

\begin{figure}[b]
	\includegraphics[width=1\columnwidth]{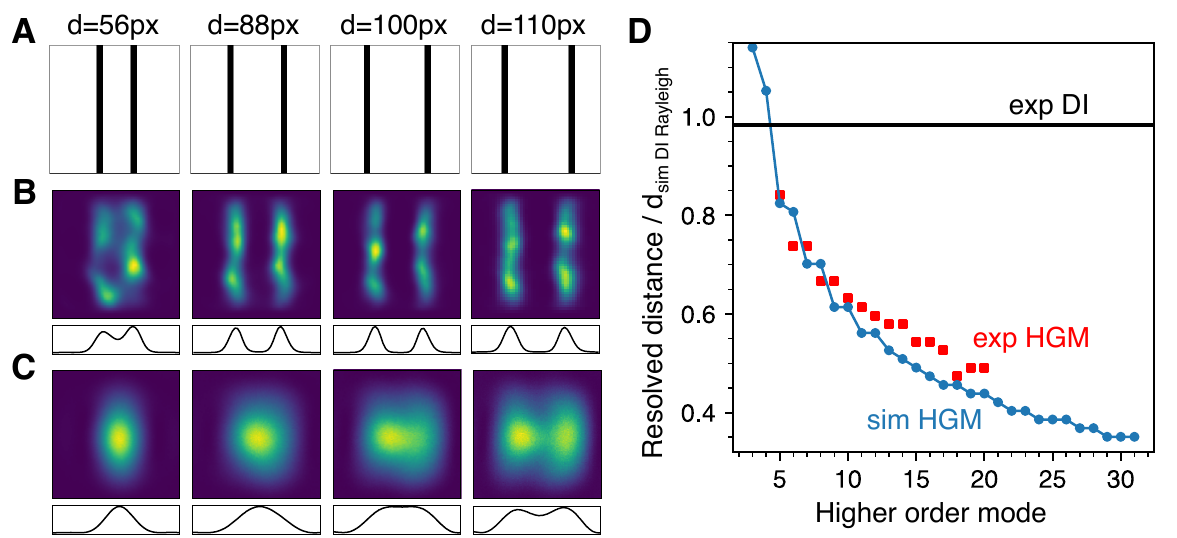}
	\caption{Qualitative assessment of the resolution improvements of HGM.
	(A) Lines pairs of increasing spacing, (B) their reconstruction via HGM using $21\times21$ modes and (C) camera images.
	(D) Rayleigh criterion for experimental (red squares) and \rednew{simulated} (blue connected dots) HGM vs number of used modes, normalized to the value for simulated direct imaging, together with the experimental direct imaging (continuous black line).
	}
	\label{fig:rayleigh}
\end{figure}

The complex photocurrent values associated with the 441 LO modes constitute the experimental data for the image reconstruction. Note that both the amplitudes and phases are needed for the imaging process: without the phases, we are unable to reconstruct the antisymmetric features of the images \cite{yang2016far}.

The HGM image reconstruction as described in Ref.~\cite{yang2016far} relies on precise knowledge of the point spread function and is supremely sensitive to even the slightest experimental imperfections. The sources of errors can be manifold: imperfect HG modes, phase aberrations in both beams' paths, the intrinsic curvature of the DMD surface,  hardness and asymmetry of the aperture, and the misalignment between the HG modes and the signal beam, among others. To overcome these issues, we calibrate our imaging system using a supervised feedforward NN \cite{palmieri2020neural}, schematized in Fig.~\ref{fig:NN}. The input of the network is 441 real and imaginary components of the heterodyne output photocurrents; the output is a 50$\times$50 bitmap containing the image. The NN architecture is shown in Fig.~\ref{fig:NN} and contains two hidden layers with 6000 units each.

In order to train the NN, we use the DMD to produce 26501 training samples consisting of random bitmaps as well as simple geometric shapes (see Supplementary Materials). These are divided into the training and cross-validation dataset in the 90:10 proportion. The  elements of the training set are sequentially displayed on the DMD, and the corresponding set of complex-valued photocurrents is acquired for each of them. 

For the corresponding  set of training labels, we do not use the ``ground truth" objects  (as done e.g.~in Ref.~\cite{ouyang2018deeplearning}); instead, we simulate the images that would be reconstructed from the training set with the ideal HGM  \cite{yang2016far,HGM_github}. With this approach, we train the NN to approximate the underlying HGM model and filter out the experimental noise, and not to guess the sample features beyond the resolution capabilities of the optics. We found that using the ground truth objects  (or even a slightly smoothed version thereof) as labels leads to overfitting issues that actually degrade the imaging quality for the test set.


After training, we evaluate the NN performance on previously unseen samples: the logo and coat-of-arms of Oxford University [Fig.~\ref{fig:reconstructions}(a,b)], pairs of lines of varied separation [Fig.~\ref{fig:rayleigh}(a,b)] and alphabet characters [Supp.~Fig.~\ref{fig:alphabet_prediction}(a,b)]. The logo and coat-of-arms have been split into respectively 30 and 120  smaller square rasters of size $210\times 210$ DMD pixels (shown by the grid in Fig.~\ref{fig:reconstructions}), each displayed on the DMD and fed to the NN in sequence. This procedure is equivalent to the transverse scanning of the object with the stride of $210\times 7.56 = 1588\ \mu$m.

In order to compare our performance with traditional direct imaging methods, we perform direct intensity measurements with a camera placed at the image plane of the objective lens [Figs.~\ref{fig:reconstructions}(c) and \ref{fig:rayleigh}(c);  Supp.~Fig.~\ref{fig:alphabet_prediction}(c)]. For the logos, the transverse scan has also been simulated as described above, albeit with a smaller stride (10 DMD pixels) because direct imaging was more sensitive to aberrations. Qualitatively, we can see that the HGM reconstructions are much sharper than direct images and allow us to see fine details and features which otherwise could not be distinguished. HGM is also superior to camera images post-processed with deconvolution algorithms (see Supplementary Materials).

In Fig.~\ref{fig:rayleigh}, we quantitatively benchmark the resolution gain by imaging pairs of parallel lines. Examples are shown in Fig.~\ref{fig:rayleigh}(a-c), whereas Fig.~\ref{fig:rayleigh}(d) plots the HGM resolution as a function of the number of HG modes used in both dimensions. To quantify the resolution, the classic Rayleigh  criterion is used, i.e. that two sources are considered resolved when the intensity at their midpoint is at most $75 \%$ of the maximum intensity. We find that HGM with up to the $HG_{20,20}$ mode can resolve two sources at approximately one-half of the the diffraction limit. In other words, the HGM resolution is comparable to the direct imaging performed using a lens that is twice as wide. 

We can also see from Fig.~\ref{fig:rayleigh}(d)  that our experimental results on the resolution are close to the \rednew{simulations}. The HGM resolution is expected to scale approximately as the inverse square root of the number of modes in each dimension. Hence the theoretically achievable resolution enhancement is significantly higher than that shown here. \rednew{In practice, an important  limitation is associated with generating high-order modes, which is increasingly challenging to do} with high fidelity due to the limited SLM resolution.
    
\rednew{Superresolution is known to be dramatically degraded by noise \cite{oh2021noisy}.} The ultimate resolution limit arises from the shot noise, \rednew{which affects the signal from all modes, but especially}  high-order HG modes, whose magnitudes rapidly fall with the mode number \cite{yang2017fisher,lupo2020noisy}. \rednew{
This is not a limiting factor in our experiment, as the number of photons in each measurement is on a scale of $10^{10}$, corresponding to the signal-to-shot-noise ratio of 50 dB. However, we simulated the shot noise effect in the Supplementary Material and found that, in the presence of that noise, HGM consistently produces higher quality reconstruction than direct imaging, yielding reasonable reconstruction quality with as few as $10^3$ photons per image section.}
 
 We now briefly discuss the perspectives of adapting our method for practical imaging, e.g. in microsopy or astronomy. One important difference is that the light sources in practical settings are typically incoherent. In this case, the phases of the heterodyne detector photocurrents are random, but their amplitudes are sufficient to reconstruct the component of the image that is symmetric with respect to the reflection about the horizontal and vertical coordinate axes \cite{yang2016far}. The  antisymmetric components can then be reconstructed by using superpositions of HG modes as the LO \cite{tsang2017subdiffraction}, or obviated by shifting the object to a single quadrant of the reference frame \cite{yang2016far}. 

A further limitation of heterodyne detection is the detectable bandwidth, which is bounded by the detector electronics. For practical imaging of broadband objects, one of the spatial mode demultiplexing methods \cite{zhou2018hermite,hiekkamaki2019perfect,fontaine2018sorter} must instead be used.



For the NN training in a microscopic setting, one could rely on off-the-shelf calibration slides and microplates for optical microscopes \cite{slide_calibration, argolight}. A calibration slide of a few tens of $\mu$m size, containing several thousand training objects of size 0.5--1 $\mu$m, can be fabricated with a resolution of a few tens of nanometers by way of lithography or laser writing. This slide can be scanned in front of the microscope objective in various orientations to increase the straining set size. If the variety and amount of the training data are still not sufficient, one could implement data augmentation techniques \cite{data_augmentation_2019}, adapting them to the HGM NN. 

HGM is a vastly simpler and cheaper alternative to many existing super-resolution methods. Furthermore, its passive nature permits universal application in a wide variety of imaging scenarios, including those not accessible by existing schemes. HGM can be combined with other imaging techniques to further increase the resolution \cite{bearne2021}. This could open up a whole new direction in both the academic and industrial sides of optical imaging.

\bibliography{scibib}

\onecolumngrid


\section*{Supplementary materials}


\subsection{Optical setup}
\label{MM:optical_setup}

The experimental setup is displayed in Fig.~\ref{fig:setup}. The beam of a continuous wave diode laser (Eagleyard EYP-DFB-0785), operating at 785 nm, is sent through a single-mode fibre (Thorlabs P3-780PM-FC-1) in order to obtain Gaussian spatial profile. A half-wave plate (HWP) and a polarizing beam splitter (PBS) further split it into two paths, which will correspond to the local oscillator (LO) and signal of the heterodyne detector (HD).

In the LO path, the laser beam is magnified by a telescope ($f_1$ = 50 mm, $f_2$ = 200 mm) in order to fully and (almost) uniformly illuminate the screen of a reflective phase-only liquid-crystal-on-silicon SLM. The incident beam hits the SLM screen almost perpendicularly, so that the angle between the incoming and reflected waves is smaller than 5 degrees, and polarized parallel to the SLM's LC director axis, so that all the incident light is modulated. The SLM holograms output the desired phase and amplitude profiles in the first order of diffraction, which is selected by a telescope ($f_1$ = 250 mm, $f_2$ = 100 mm) and an iris at its focal plane.

In the signal path, the beam is sent through an acousto-optics modulator (AOM, Isomet 1205C-1), driven at 92.05 MHz. The produced first diffraction order mode is incident upon a DMD (DLP LightCrafter 6500), which modulates its amplitude with the binary bitmap of the object to be imaged. The imaging system aperture is set by an iris whose diameter is set to $3.5\pm0.03$ mm and placed at a distance of $245.5\pm1$ cm from the DMD, corresponding to an optical system of numerical aperture (NA) of $(7.1 \pm 0.07) \times$ 10$^{-4}$. The iris diameter was measured by placing a metal ruler next to it and imaging both these objects with a high-resolution camera.



After the iris, a set of three telescopes magnify and collimate the signal beam in order to match the waist and wave-vector of the LO beam in the 0th order HG mode. The first telescope ($f_1$ = 100 mm, $f_2$ = 50 mm) collects the light transmitted through the iris. In the second ($f_1$ = 50 mm, $f_2$ = 100 mm) and third telescopes ($f_1$ = 75 mm, $f_2$ = 75 mm), the ``eyepiece" lenses are mounted on translation stages to independently control the signal beam’s diameter and divergence, hence enabling mode matching to the LO. The signal and LO paths are reunited at a PBS, whose output beams feed the photodetectors of a homemade balanced detector \cite{kumar2012homodyne,masalov2017}.

\subsection{Spatial light modulator}
\label{MM:SLM}

The SLM (Hamamatsu X13138-02, 1272$\times$1024 pixel resolution and 12.5 $\mu \textup{m}$ pixel pitch) modulates the phase of an incoming optical wave by controlling the effective refractive index of the liquid crystal layer in each of its pixels. In order to generate  Hermite-Gaussian modes, we display a phase grating (``hologram") on the SLM screen. Setting the grating’s depth and offset at each point of the SLM surface allows us to generate any desired complex spatial profile in the first order of diffraction in the reflected wave \cite{bolduc2013exact}. 

\begin{figure}[h]
	\includegraphics[width=0.8\columnwidth]{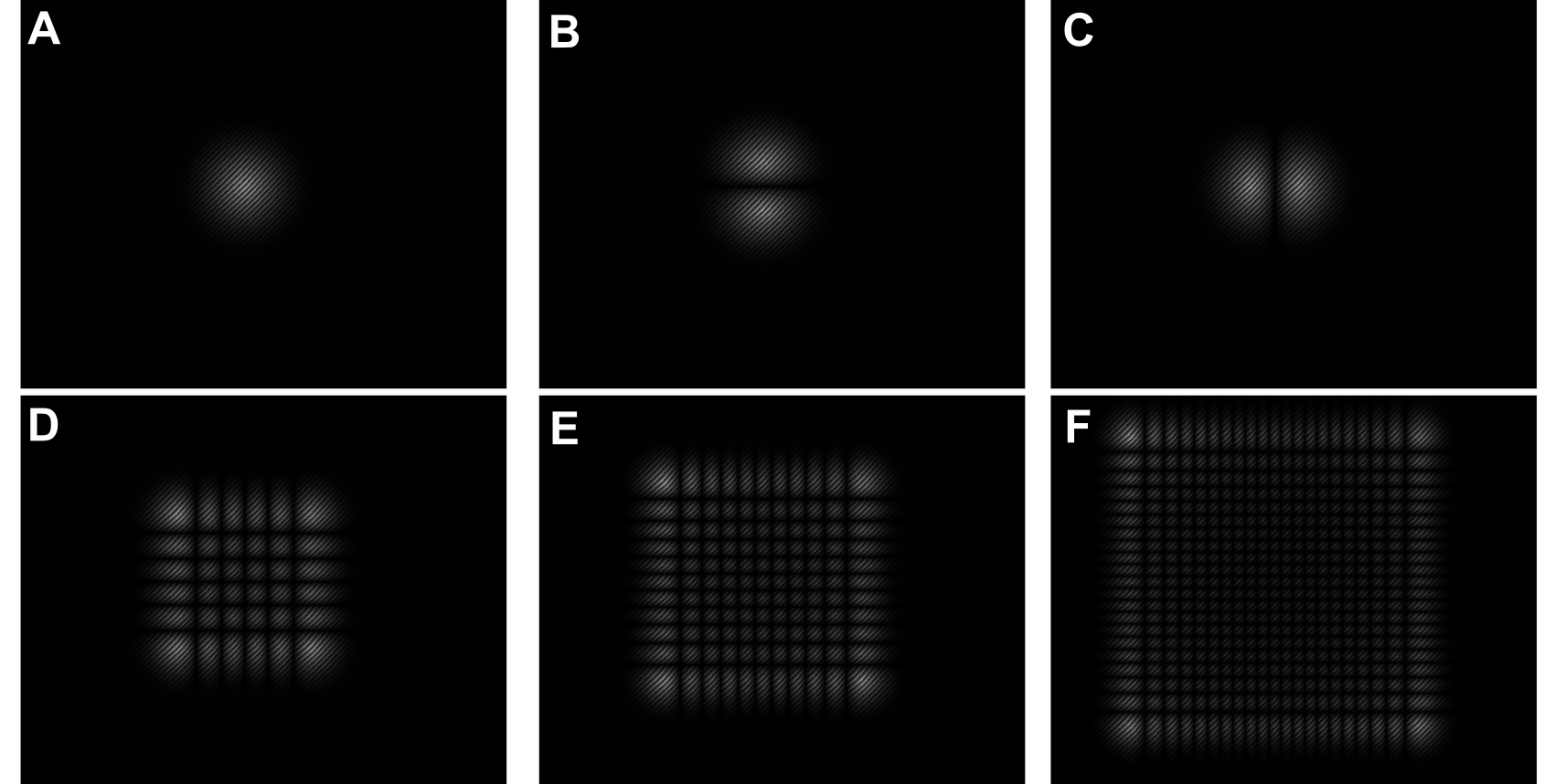}
	\caption{Examples of SLM holograms used to produce HG modes.
		(A) $HG_{0,0}$, (B) $HG_{0,1}$, (C) $HG_{1,0}$,  (D) $HG_{5,5}$, (E) $HG_{10,10}$, (F) $HG_{20,20}$.
	}
	\label{fig:holograms}
\end{figure}

Fig.~\ref{fig:holograms} illustrates a few examples of the resulting holograms. When computing the hologram, we compensate for the incoming beam's non-uniform profile and the SLM backplane curvature using the procedure described in Ref.~\cite{pushkina2020slm}. The holograms can be modified further to  alter the waist, displacement and orientation of the generated modes. 

\subsection{Digital micromirror device and generated sources}
\label{MM:DMD}
\indent
The DMD consists of 1920$\times$1080 micromirrors. Each of them can be set to the ON or OFF state, corresponding, respectively, to a tilt by $+12^\circ$ or $-12^\circ$ along their diagonal axis and relative to their flat position. The signal beam illuminates the central area of the DMD, where 210$\times$210 micromirrors display the objects to be imaged. 
The DMD micromirrors outside that area are set to the OFF state. The micromirror pitch is 7.56 $\mu \textup{m}$, so the total working area is 1.588 mm wide.

The DMD is driven in ‘pattern on-the-fly’ mode, in which a sequence of 400 binary bitmap images is loaded into the internal memory of the DMD controller board via USB. Each loaded sequence includes 398 objects to be imaged, a phase reference object and an alignment
square, whose purpose is described below. 

\begin{figure}[h]
	\includegraphics[width=0.6\columnwidth]{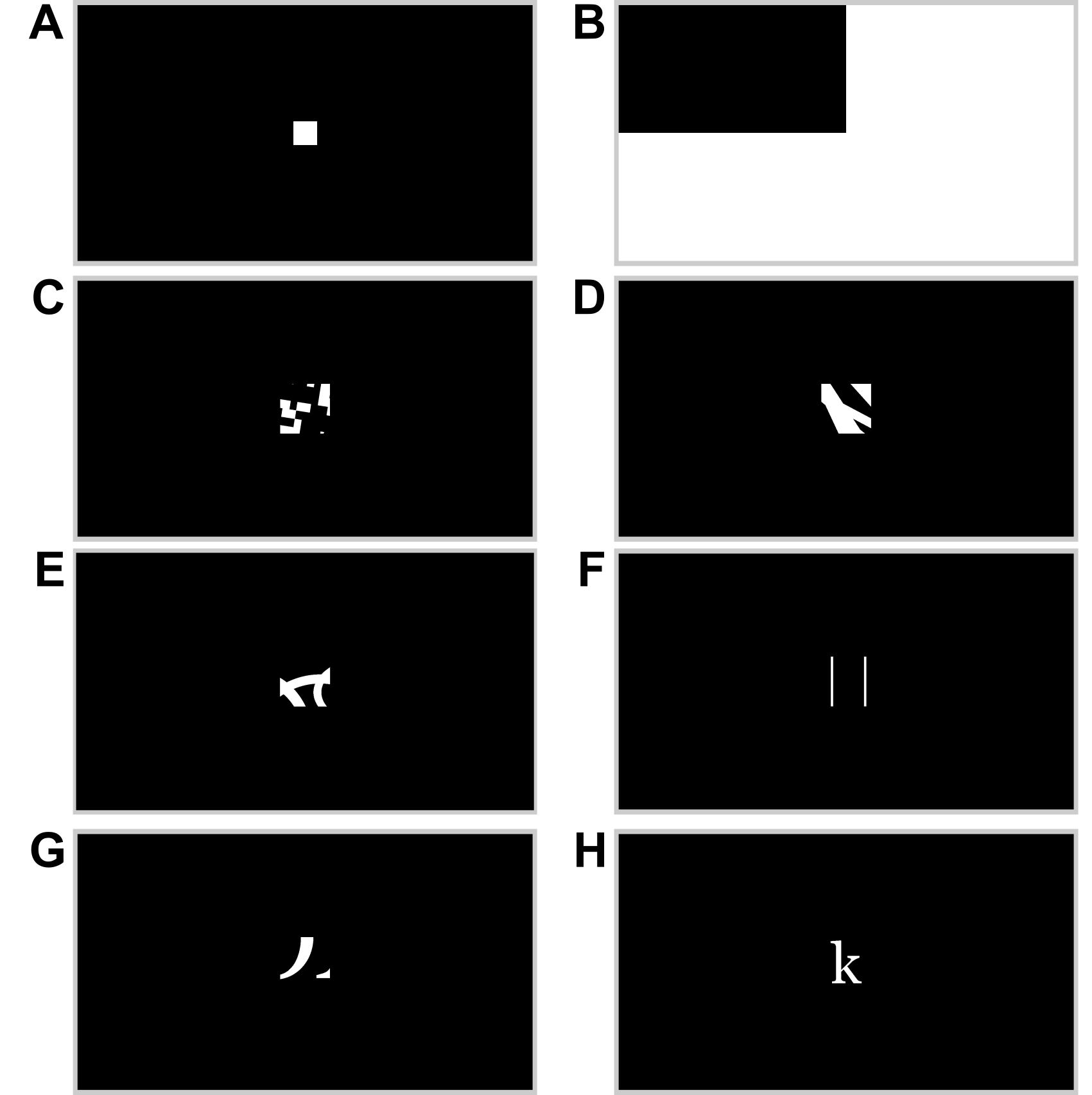}
	\caption{Examples of DMD frames used in the experiment: (A) alignment square; (B) phase reference frame; (C) random matrix (train dataset); (D) superposed random lines (train dataset); (E) random parts of an ellipse (train dataset); (F) pair of lines (test dataset); (G) frame from Oxford University logo (test dataset); (H) text symbol (test dataset).
	}
	\label{fig:DMD_frames}
\end{figure}



\subsection{Acquisition}

We set the acquisition order to minimize the overall measurement time, taking into account the SLM and DMD refresh times ($\sim0.5$ s and $\sim0.1$ ms, respectively), and the time required to load new frames into the DMD internal memory ($\sim 20$ s for a batch of 400 binary frames). After each such batch is loaded into the DMD, it is displayed sequentially while keeping the LO mode constant. The corresponding 400 photocurrents are acquired with a digital oscilloscope in a single trace, along with the AOM driving signal to keep track of the phase. While the trace is transferred from the
oscilloscope to the PC in the binary format via USB (it takes approximately 1 s to transfer 55000 data points and 4 channels), the SLM hologram is updated to produce the next HG mode. The acquisition for one such batch and all 441 LO modes  lasts about 9 minutes. Acquiring the training set ($26501$ DMD frames) takes about 10 hours, while the test set (Oxford logos, text symbols and line pairs) is limited to $302$ DMD frames and is hence acquired as a single batch.

The acquired traces are processed to extract the amplitude and phase of the heterodyne detection photocurrent for each object. The relative phase between the signal and LO arms of the optical setup drifts with time. To keep track of this drift, we use the phase reference object [Fig.~\ref{fig:DMD_frames}(B)] that is the union of the bottom and top-right quadrants of the DMD screen. Its asymmetric shape has nonzero overlap with all LO modes, so this object always generates a measurable photocurrent. The phase $\phi_{j,m,n}$ associated with the LO mode HG$_{m,n}$ and object $j$ can then be calculated according to
\begin{equation}
	\phi_{j,m,n} = \left(\phi_{j,m,n}^{\rm HD} - \phi_{j,m,n}^{\rm AOM} \right) - \left(\phi_{{\rm ref},m,n}^{\rm HD} - \phi_{{\rm ref},m,n}^{\rm AOM}\right),
\end{equation}
where each term is parentheses is the difference between the phases of the photocurrent and the AOM driving signal acquired by the oscilloscope, both oscillating at 92.05 MHz. 

The Python code used for the acquisition and pre-processing can be seen in our GitHub repository \cite{HGM_github}.

\subsection{Automatic interferometer alignment procedure}
\indent
Because the data acquisition takes several hours, it is necessary to regularly realign the LO and signal beams with respect to each other in both transverse position and direction. To this end, we have the DMD display at its centre the alignment square of size 100$\times$100 pixels [Fig.~\ref{fig:DMD_frames}(A)]. When the signal and LO are perfectly aligned, we expect null photocurrents when the LO is in modes HG$_{1,0}$ and HG$_{0,1}$. The vertical or horizontal misalignment will give rise to a nonzero photocurrent for the corresponding LO mode.
This error signal is reduced by modifying the displacement and phase gradient of the SLM hologram. The procedure lasts for about 3 minutes and is performed every 3 hours. The corresponding code can be found in Ref.~\cite{HGM_github}.

Note that the alignment square is also displayed in every batch of objects during the regular data acquisition for monitoring purposes.

\subsection{Neural Network}




The NN architecture is shown in Fig.~\ref{fig:NN} and consists of two hidden layers with 6000 units each. Since the normalized input ranges from $-1$ to $1$, we use the hyperbolic tangent as activation function for the first layer. For the second layer, we use a ReLU activation for train speed purposes and, after the third, a sigmoid, in order to adapt to the $[0,1]$ range of the labels. The optimization method is Adaptive Moment Estimation (Adam), with a learning rate of $10^{-4}$, exponential decay rate moving parameters for the first and second moment estimates $(0.9, 0.999)$ and weight decay $0$.
We set the batch size to $512$. We notice that a large batch size, apart from reducing the training time, allows achieving lower training and cross-validation losses. 
The loss function is the mean squared error (MSE) loss, which makes the NN effectively behave as a nonlinear regressor. The NN is trained for 900 epochs, achieving training and validation log$_{10}$-loss of $-6.674$ and $-6.498$ per sample, respectively, as illustrated in the inset of Fig.~\ref{fig:NN}.

\subsection{Train, validation and test datasets}
The training/cross-validation dataset consists of
\begin{itemize}
	\item 20000 frames that are tiled with black and white rectangles. The tile size is randomly chosen from 10 to 50 pixels in each dimension and the ``color" of each tile is also random, sampled from a uniform distribution between 0.2 and 0.8; the entire matrix is then randomly rolled along both axes and randomly oriented, Fig.~\ref{fig:DMD_frames}C;
	\item 3000 random sets of 1 to 5 lines with random orientation, position and width ($10$ to $70$ pixel), Fig.~\ref{fig:DMD_frames}D;
	\item 3000 random sets of 1 to 5 segments of elliptic rings with randomly variable eccentricity and width (minor and major axis from $40$ to $120$ pixel, width from $10$ to $15$ pixel), Fig.~\ref{fig:DMD_frames}E;
	\item 58 single squares of sizes $30$ pixel and $70$ pixel at various positions within the  $210 \times 210$ frame;
	\item 441 squares of $30$ pixel size at variable positions within the $210 \times 210$ frame, from which a smaller square of $10$ pixel size is subtracted, with variable positions of the smaller square with respect to the larger one;
	\item a blank image;
	\item a square occupying the entire $210 \times 210$ pixel frame.
\end{itemize}
During the training time, the train/cross-validation dataset is randomly split into 90\% for training and 10\% for validation.

\begin{figure}[h]
	\includegraphics[width=1\columnwidth]{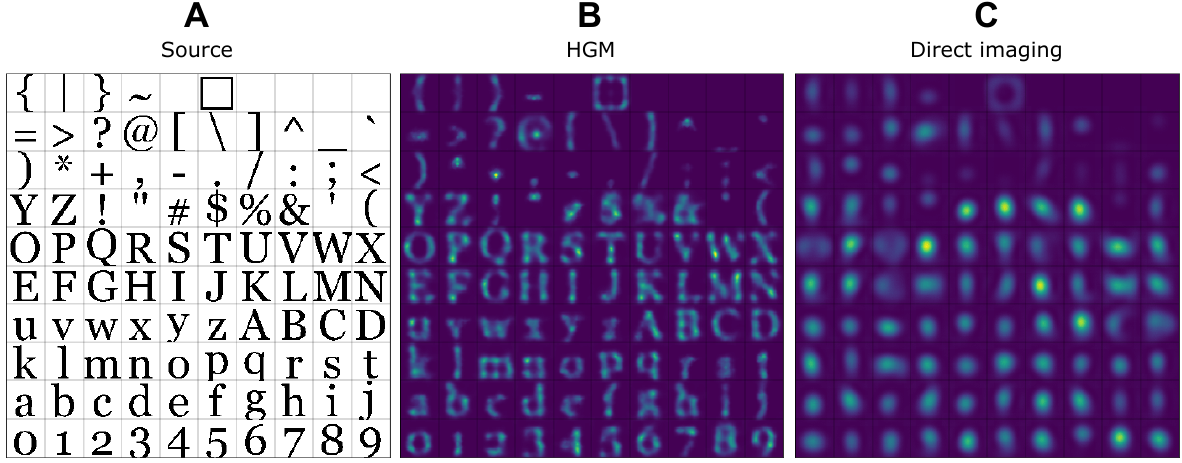}
	\caption{Original (A), HGM-reconstructed (B) and camera (C) images of the \textit{text symbols} test set.}
	\label{fig:alphabet_prediction}
\end{figure}

The test dataset consists of
\begin{itemize}
    \item University of Oxford logo, divided into 10$\times$3 frames [Fig.~\ref{fig:reconstructions}(A), top].
    \item University of Oxford coat of arms, divided into 10$\times$12 frames [Fig.~\ref{fig:reconstructions}(A), bottom].
    \item 96 text symbols from the Latin alphabet and special ASCII characters [Fig.~\ref{fig:alphabet_prediction}(A)].
    \item 56 centred pairs of vertical lines of $10$ pixel thickness and varied spacing, used to evaluate the Rayleigh distance.
\end{itemize}

These datasets can be downloaded at \cite{HGM_github}.

\subsection{Direct imaging experiment and deconvolution algorithm}

In order to assess the resolution improvement of HGM with respect to conventional microscopy, we perform direct intensity measurement of the image field with a CMOS camera. For the parallel lines and alphanumerical symbols, a single direct image of the $210\times210$ pixel object is acquired. For the two larger objects, we simulate the scanning as described in the main text. 

To further evaluate the performance of HGM, we post-process the experimental camera images using one of the commonly used deconvolution algorithms,
Richardson-Lucy (RL) deconvolution. RL is an iterative algorithm that aims to estimate the original object $O(x,y)$, given the point spread function $P(x,y)$ and the experimental camera image $I(x,y)$, via
\begin{equation}
\hat{O}_{t+1} =\hat{O}_{t} \cdot \Bigg( \frac{I}{\hat{O}_{t} * P} * P \Bigg) 
\end{equation} 
with $\hat{O}_t$ being the estimate of the object at the  iteration $t$ and asterisks denote convolution. 

\begin{figure}[h!]
	\includegraphics[width=1\columnwidth]{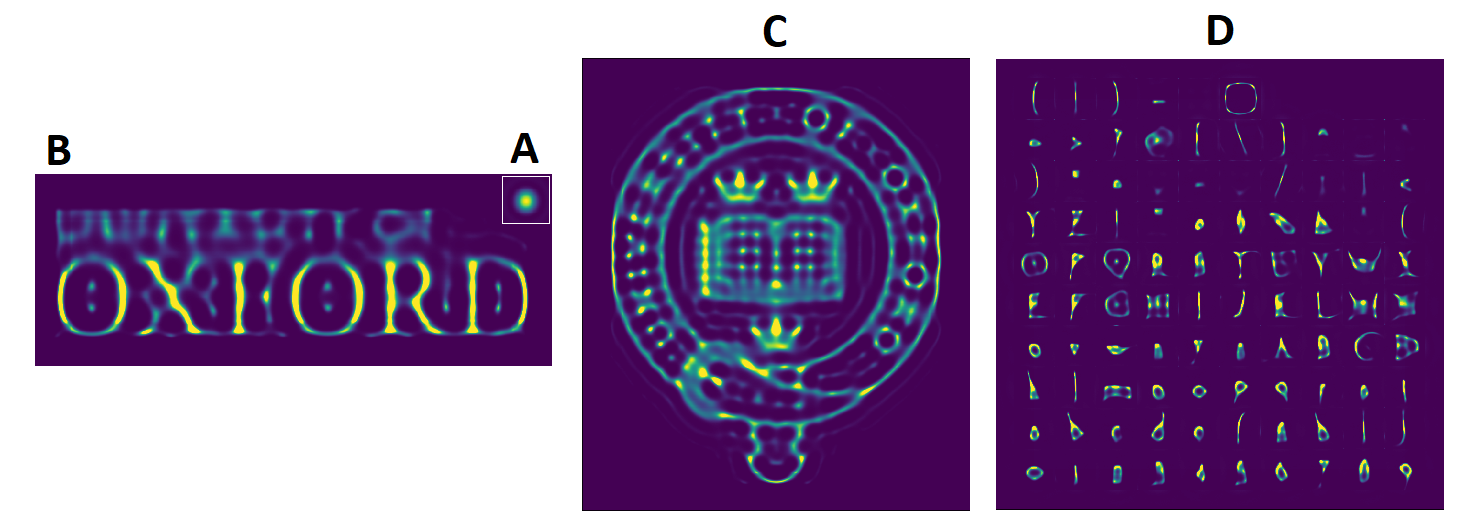}
	\caption{RL applied to the camera images of the test datasets using the Airy point spread function (A). Estimates of the two University of Oxford logos (B,C) and text symbols (D).}
	\label{fig:RL}
\end{figure}

Fig.~\ref{fig:RL}(B-D) shows the result of RL applied to the test datasets (two University of Oxford logos and the text symbols test), after $1000$ iterations and using the Airy point spread function (PSF) reported in Fig.~\ref{fig:RL}A. The size of the Airy PSF is chosen to match the experimental PSF, with 45 camera pixels from its centre to the radius of the first zero. The program used to perform RL can be found in \cite{HGM_github}.

By qualitatively comparing the HGM predictions (Fig.~\ref{fig:reconstructions} and Fig.~\ref{fig:alphabet_prediction}) with the results of RL (Fig.~\ref{fig:RL}), we conclude that HGM allows for better reconstruction of the original object, with a higher resolution and fewer artifacts.


\subsection{Alternative resolution improvement benchmark}


\begin{figure}
	\includegraphics[width=1\columnwidth]{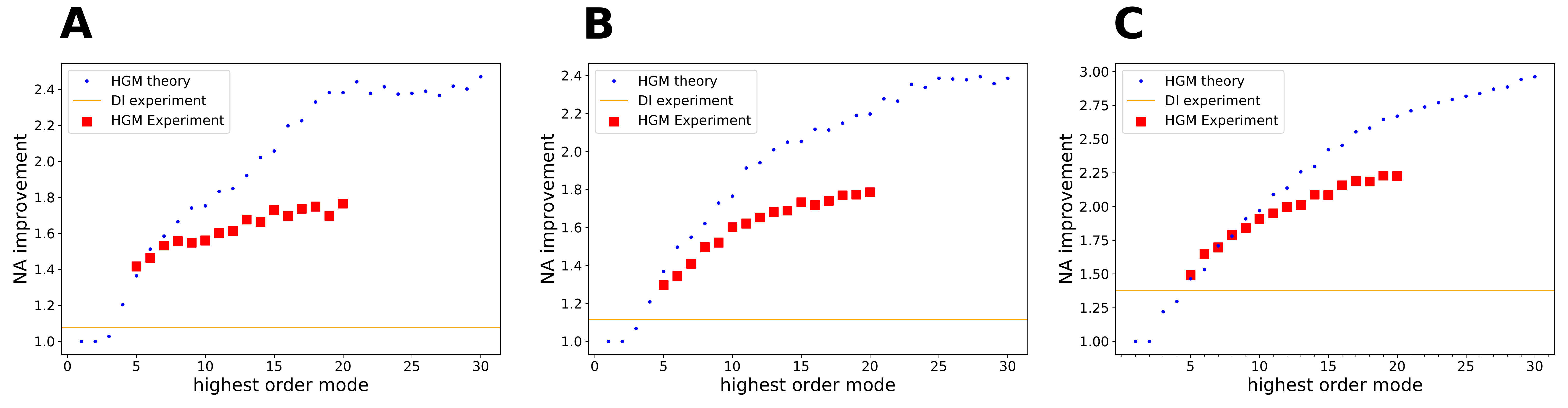}
	\caption{Resolution improvement for Oxford logo, Oxford coat of arms and and the text symbols test set, evaluated through the MSE loss as described in the text. For the 20$\times$20 reconstructions, we achieved resolution improvements of $1.76$, $1.78$ and $2.22$, respectively.}
	\label{fig:MSE}
\end{figure}

An alternative way of estimating the resolution improvement is by evaluating the mean squared error (MSE) between the reconstructed images and the original objects. We calculate the MSE $M_{\rm DI}({\rm NA})$ between theoretically calculated direct camera images with varying numerical apertures and the original objects, and then compare these values with the MSE $M_{\rm HGM}({\rm NA}_0)$ for the HGM reconstruction with the numerical aperture ${\rm NA}_0$. Because HGM is generally superior to direct imaging, $M_{\rm DI}({\rm NA}_0)>M_{\rm HGM}({\rm NA}_0)$. However, the function $M_{\rm DI}({\rm NA})$ monotonically decreases with ${\rm NA}$ and becomes equal to $M_{\rm HGM}({\rm NA}_0)$ at some ${\rm NA}_1>{\rm NA}_0$: in other words, the image quality obtained via direct imaging with the numerical aperture ${\rm NA}_1$ is similar to that with HGM and numerical aperture ${\rm NA}_0$. The quantity ${\rm NA}_1/{\rm NA}_0$ corresponds to the resolution enhancement and is plotted, as a function of the number of modes used in HGM, in Fig.~\ref{fig:MSE}. 

\rednew{\subsection{Hermite-Gaussian microscopy and shot noise}}

In this section we recapitulate the derivation of the HGM equations for coherent light as in Ref.~\cite{yang2016far} and extend it for two-dimensional imaging. We also discuss how HGM behaves in noisy environments.

Consider a plane object illuminated by a uniform coherent light beam and imaged by an objective lens. At the image plane of the lens, we have the field distribution
\begin{equation}
	E'(x,y) = \int \int_{-\infty}^{+\infty} E(x',y') \psi(x-x',y-y') dx'dy',
\end{equation}
where $(x,y)$ and $(x',y')$ are the image and object plane coordinates, respectively, $E(x,y)$ is the object plane field distribution and $\psi(x,y)$ the point-spread function (PSF) of the lens. This PSF can be well approximated by a ``soft-aperture'' Gaussian PSF $\psi(x,y) = e^{-(x^2+y^2)/4\sigma^2}/\sqrt{2 \pi \sigma^2}$, where $\sigma \approx 0.21 \lambda / \textup{NA}$ \cite{zhang2007gaussian}. Instead of directly detecting the light's intensity $|E'(x,y)|^2$, we perform heterodyne detection using as LO the $(n,m)$-th order HG mode 
\begin{equation}
	\psi_{mn}(x,y) = \frac{1}{2^n n!}  H_n\left(\frac{x}{\sqrt{2}\sigma}\right) H_m\left(\frac{y}{\sqrt{2}\sigma}\right)\psi(x,y),
\end{equation}
with the same waist $\sigma$ as the PSF. The resulting heterodyne photocurrents are then proportional to the overlap between the signal and LO beams,
\begin{equation}\label{eq:HGM_J_n}
	J_{mn} \propto \int \int_{-\infty}^{+\infty}E'(x,y)\psi_{mn}(x,y) dx dy = \frac{1}{\sqrt{m! n!}}\int \int_{-\infty}^{+\infty}E(x,y)\left(\frac{x}{2 \sigma}\right)^m \left(\frac{y}{2 \sigma}\right)^n e^{-(x^2+y^2)/8\sigma^2} dx dy.
\end{equation}

The original object plane field can be recovered from the measured photocurrents by calculating the coefficients
\begin{equation}\label{eq:HGM_beta}
	\beta_{mn} = \sum_{k=0}^{m}\sum_{l=0}^{n}\sqrt{m! n!} \alpha_{mk}\alpha_{nl} J_{kl}, 
\end{equation}
where the $\alpha_{mk}$ are the coefficients of Hermite polynomials, $H_m(x) = \sum_{k=0}^{m} \alpha_{mk} x^k$, and then reconstruct the object field $E(x,y)$ via
\begin{equation}\label{eq:HGM_reconstruction}
	E_{\textup{rec}}(x,y) = \sum_{m,n=0}^{\infty}  \frac{\beta_{mn} H_m\left(\frac{x}{2\sigma}\right)H_n\left(\frac{y}{2\sigma}\right)e^{-(x^2+y^2)/8\sigma^2}}{2^{m+n+2}m!n!\pi \sigma^2}.
\end{equation}

\begin{figure}
	\includegraphics[width=1\columnwidth]{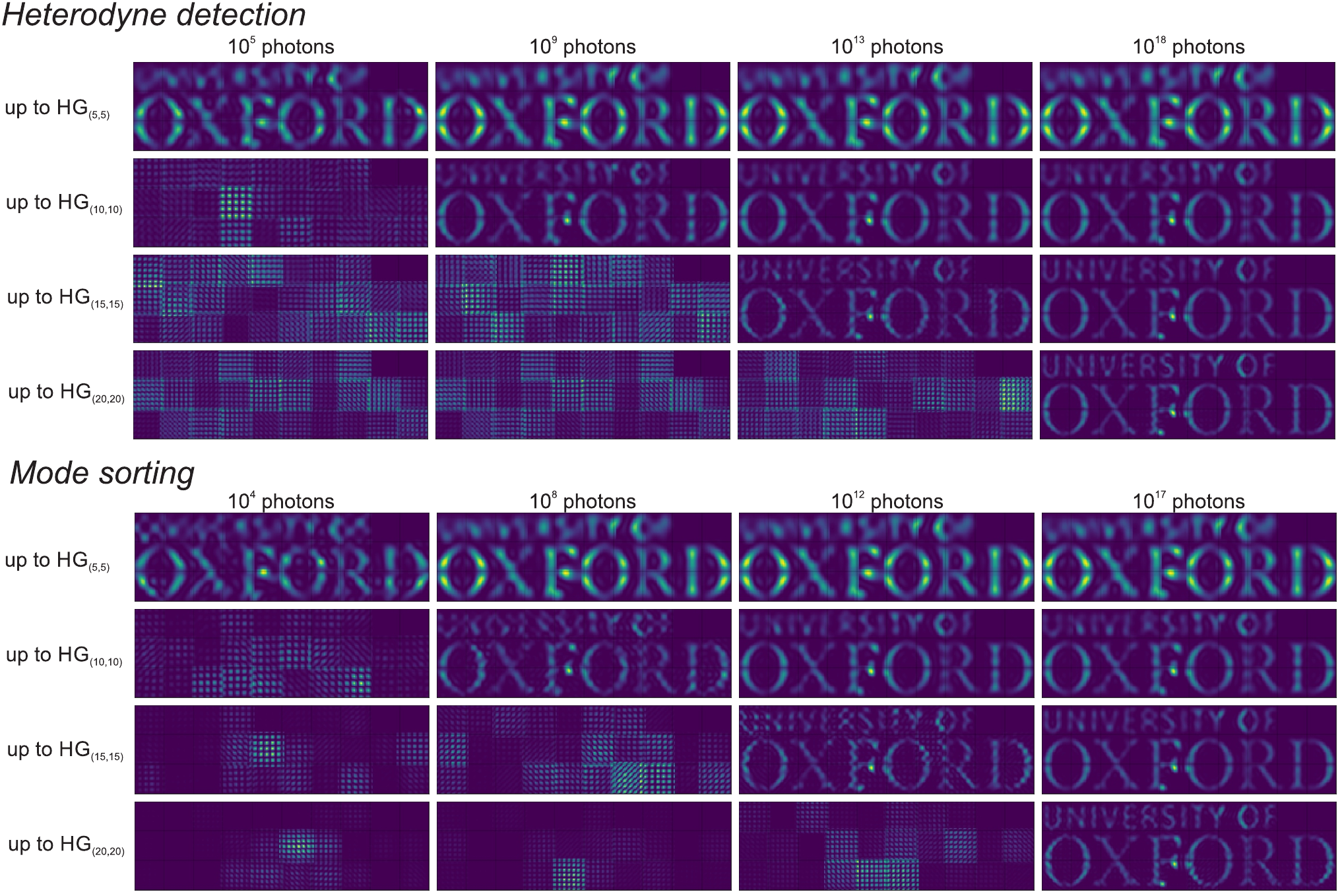}
	\caption{Oxford logo reconstruction using the closed-form expression \eqref{eq:HGM_reconstruction} in the presence of shot noise. Simulations for two detection methods, different sizes of HG mode basis and different number of photons available per image section (DMD frame) are displayed. When fewer photons are available, the relative shot noise is more significant and reasonable reconstruction is possible only with smaller HG bases. }
	\label{fig:new_HGM}
\end{figure}

These equations are derived in a noiseless, ideal scenario. In a real experiment, even if all sources of systematic errors are eliminated, the unavoidable shot noise will still be present. As can be seen from Eq.~\eqref{eq:HGM_J_n}, the photocurrents get exponentially smaller with the HG mode order, and therefore become heavily affected by shot noise. These small photocurrents, however, are  amplified via the factorial terms in Eq.~\eqref{eq:HGM_beta}. The result is that even the slightest noise leads to a complete breakdown of the reconstruction in Eq.~\eqref{eq:HGM_reconstruction}

In Fig.~\ref{fig:new_HGM} we simulated an idealized scenario where there are no experimental imperfections and the only source of error is shot noise (which is the inverse square root of the number of photons available). The results show that $10^{17}$--$10^{18}$ photons per image section are necessary for HGM to work when HG modes up to (20,20) modes are used, which is an enormous photon overhead. Even when limiting the mode order to  (5,5), still $10^4$--$10^5$ photons are necessary to produce any meaningful result.

\begin{figure}
	\includegraphics[width=1\columnwidth]{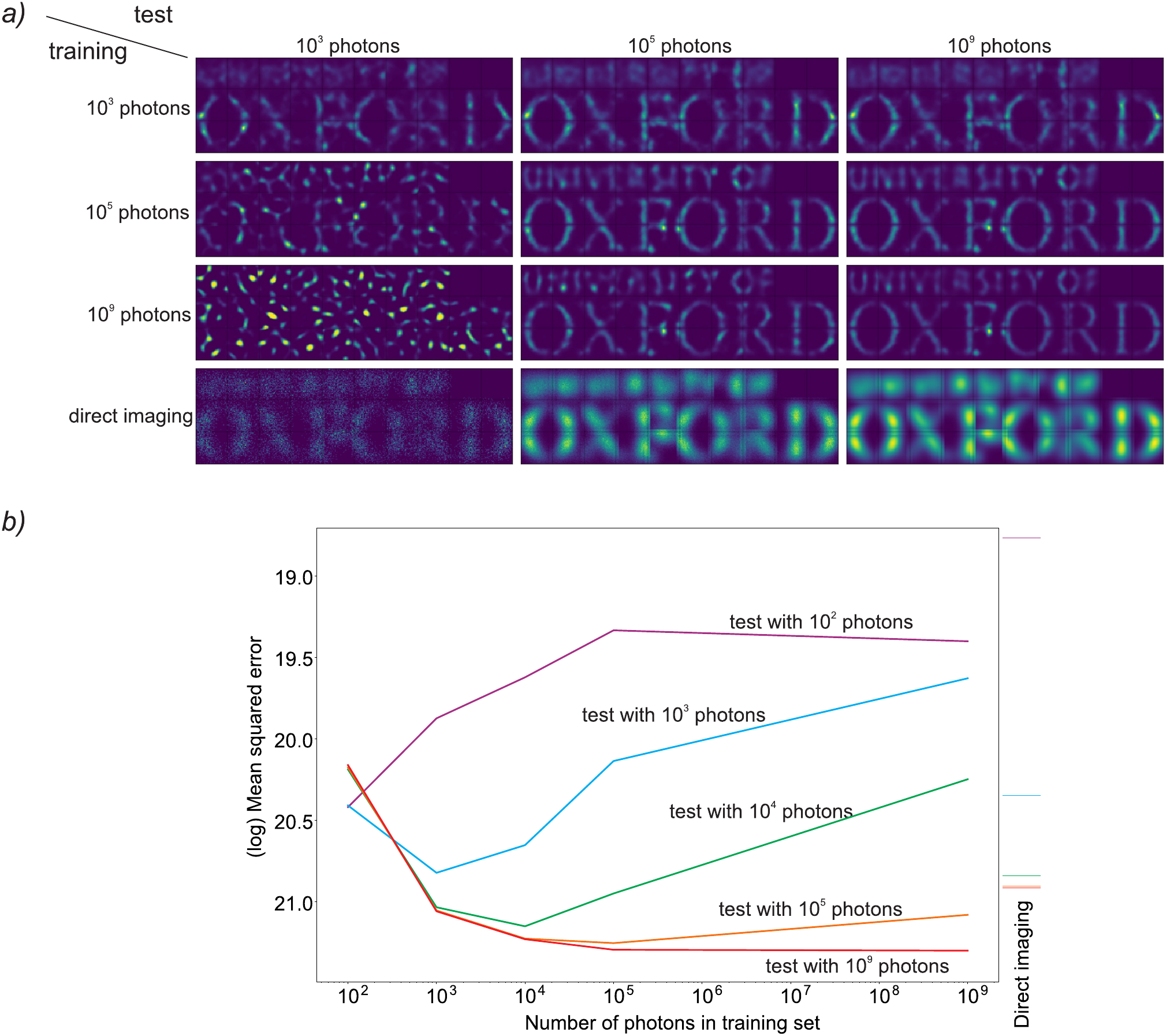}
	\caption{Simulated Oxford logo reconstruction via direct imaging and HGM using heterodyne detection with different photon numbers available per image section during training and test phases. a) Examples. b) (Logarithm of) mean squared error between the noisy reconstructed logo and the original object. Horizontal lines on the right correspond to the error obtained for the direct imaging reconstructions with the same photon numbers.}
	\label{fig:new_predictions}
\end{figure}

Therefore, practical HGM requires an algorithm that regularizes the noise from the higher-order photocurrents while at the same time being able to extract information from them. In this work, this task is addressed by the neural network, and in Fig.~\ref{fig:new_predictions} we assessed its efficiency against shot noise. We observe that although the reconstruction quality degrades with the falling number of photons, we are still able to obtain meaningful results even with a very limited number of photons. Remarkably, HGM with as low as $10^3$ photons per image section yields the same reconstruction quality as direct imaging in the absence of the shot noise. 

In Fig.~\ref{fig:new_predictions}, we also analyze how the reconstruction quality is affected when the NN trained on noisy data with certain  number of photons available (and hence a certain shot noise level) is used to reconstruct an image from data acquired with a different number of photons. For a given number of photons in the test set, the optimal reconstruction quality is obtained when training set has been acquired with the same number of photons (i.e. has the same noise level). If difference is present, the reconstruction quality is affected less significantly if the test set has a better signal-to-noise ratio than the training set, rather than the other way around. 

\end{document}